\documentstyle[epsfig,twoside,fleqn,espcrc2]{article}
\def    \be             {\begin{equation}}
\def    \ee             {\end{equation}}
\def    \ba             {\begin{eqnarray}}
\def    \ea             {\end{eqnarray}}
\def    \nn             {\nonumber}
\def    \=              {\;=\;}
\def    \frac           #1#2{{#1 \over #2}}
\def    \rd             {{\mathrm d}}    
\def\mufrag{\mbox{$\mu_{\mathrm F}$}}
\def\ups {\mbox{$\Upsilon$}}

\begin{document}
\def \QQ {Q \overline Q}
\def \qq {\mbox{$q \overline q$}}
\def    \q              {\ifmmode {\cal{Q}} \else ${\cal{Q}}$ \fi}
\def    \as             {\mbox{$\alpha_s$}}
\def    \asb            {\mbox{$\alpha_s^{(b)}$}}
\def    \assq           {\mbox{$\alpha_s^2$}}
\def    \mur            {{\mbox{$\mu_{\rm R}$}}}
\def    \mursq          {\mbox{$\mu^2_{\rm R}$}}
\def    \mul            {{\mu_\Lambda}}
\def    \mulsq          {\mbox{$\mu^2_\Lambda$}}
\def\etah {\mbox{$^1S_0^{[8]}$}}
\def\etas {\mbox{$^1S_0^{[1]}$}}
\def\psih {\mbox{$^3S_1^{[8]}$}}
\def\psis {\mbox{$^3S_1^{[1]}$}}
\def\chizh {\mbox{$^3P_0^{[8]}$}}
\def\chioh {\mbox{$^3P_1^{[8]}$}}
\def\chith {\mbox{$^3P_2^{[8]}$}}
\def\chijh {\mbox{$^3P_J^{[8]}$}}
\def\aem{\mbox{$\alpha_{{\mathrm\tiny em}}$}}

\def    \oneSzero       {\ifmmode {^1S_0} \else $^1S_0$ \fi}
\def    \threeSone      {\ifmmode {^3S_1} \else $^3S_1$ \fi}
\def    \onePone        {\ifmmode {^1P_1} \else $^1P_1$ \fi}
\def    \threePJ        {\ifmmode {^3P_J} \else $^3P_J$ \fi}
\def    \threePzero     {\ifmmode {^3P_0} \else $^3P_0$ \fi}
\def    \threePone      {\ifmmode {^3P_1} \else $^3P_1$ \fi}
\def    \threePtwo      {\ifmmode {^3P_2} \else $^3P_2$ \fi}

\newcommand{\ttbs}{\char'134}
\newcommand{\AmS}{{\protect\the\textfont2
  A\kern-.1667em\lower.5ex\hbox{M}\kern-.125emS}}

\title{Colour Octet Effects in Quarkonium Physics}

\author{F. Maltoni\address{Dipartimento di Fisica dell'Universit\`a 
and Sez. INFN, Pisa, Italy}}%

\begin{abstract}
The importance of including colour octet contributions in describing
decay and production of quarkonium states is briefly discussed for two
cases : the radiative decays of the $\Upsilon$ and the production of
$J/\psi$ in the inclusive $B$ decays.  It is shown how information on
the non-perturbative matrix elements can be obtained by comparing the
theoretical expressions computed at next-to-leading order in $\as$ with
the experimental data.

\end{abstract}

\maketitle

\section{Introduction}

It is now about four years that the non-relativistic QCD approach
for the description of phenomena involving quarkonium production and decays,
has been introduced~\cite{bbl}. Without any doubt, Bodwin, Braaten and Lepage
have provided us with a powerful and rigorous framework where 
calculations can be undertaken both at perturbative and non-perturbative
level. Within this approach, the factorization between short-distance 
and long-distance physics allows to express physical quantities 
as inclusive decay widths or  cross sections as:
\begin{eqnarray}
{\cal Q}(H) &=& \sum_n c_n \langle {\cal O}_n(H) \rangle 
\end{eqnarray}
where in $c_n$ are encoded the high-energy modes of the theory and so 
can be calculated perturbatively in QCD, while the matrix elements (MEs)
are non-perturbative and can be either obtained by lattice 
simulation~\cite{lattice} or
by comparison with experimental data. Moreover, by means 
of the power counting rules of NRQCD, it is possible to organize the MEs
in a hierarchy based on the relative velocity $v \ll 1 $ of
the heavy-quarks in the bound state. In the end one is able to 
organize all these terms in a double expansion in $\alpha_s$ and in $v$ and
to make predictions, in principle,  at any given order of accuracy.
 
With NRQCD in hands, much of the effort of the last years (for very 
nice reviews on more recent developments see for example \cite{revs}) has been
devoted mainly to improve at NLO in $\alpha_s$ the determination of
the short distance coefficient $c_n$ for the most relevant terms
in the $v$ expansion and use the data available on
quarkonium production and decays to extract information on the
non-perturbative MEs. The aim of the game 
has been to check the cornerstone of the predictive power of the theory:
the MEs should be insensitive to the details of the hard
processes and depend in a universal way only on the bound state
physics. This allows, in principle,
to extract the values of the non-perturbative parameters from 
the experimental data on one process and 
then to use them to make predictions for others. 

In the following Sections I  will discuss  the necessity of including
the colour octet contributions at NLO to correctly predict and/or explain
experimental data for two cases of interest.

\section{ The photon spectrum in bottomonium decay}

A consistent description of the photon energy spectrum in
$\ups\to\gamma\,+X$ decay requires the inclusion of the fragmentation
components
\footnote{Emission from final state light quarks has been surprisingly
considered only recently by Catani and Hautmann in the CSM framework
(and presented here in Montpellier in the QCD94 Conference
!~\cite{Catani})} within the NRQCD factorization approach\cite{MP98}.
The differential photon decay can be expressed in terms of a
convolution between partonic kernels $C_a$ and the fragmentation
functions $D_{a\to\gamma}$:
\ba
&&\frac{\rd\Gamma}{\rd z}=C_\gamma(z) +\nn\\ 
&& \sum_{a= q, \overline
q ,g}\int_z^1 \frac{\rd x}{x} C_a(x,\mufrag)
D_{a\to\gamma}(\frac{z}{x},\mufrag) 
\ea 
where $z= E_\gamma/m_Q$ is the
rescaled energy of the photon (~$m_Q$ is the heavy quark mass~).
The first term corresponds to what is usually called
the `prompt' or 'direct' photon production 
where the photon is produced directly in the hard interaction 
while the second one  corresponds to  the long distance
fragmentation process where one of the partons fragments
and transfers a fraction of its  momentum to the photon.
%
\begin{figure}[t]
\centerline{\epsfig{file=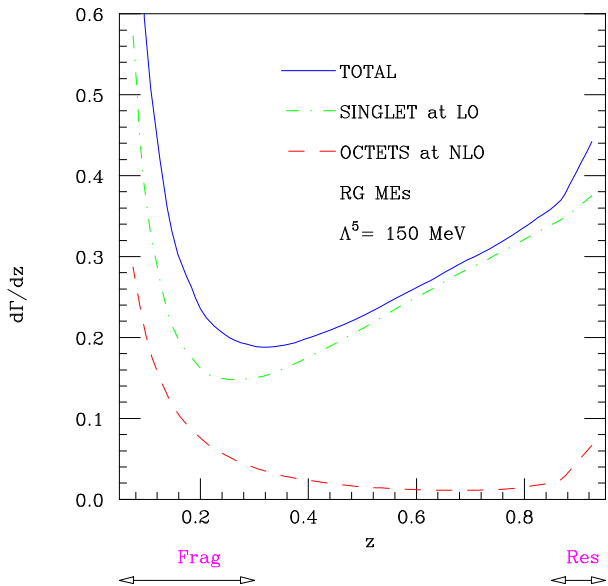,%
          height=8cm,clip=,angle=0}}
\caption[dummy]{\label{fig:plot-res2}
Total colour octet contribution on the LO, colour singlet photon
spectrum.  Notice that neither NLO QCD nor relativistic effects are
included in the singlet contribution. The overall normalization has
been chosen as to reproduce the LO result of Ref.~\cite{Catani}.}
\end{figure}
%
The NRQCD expansion for the coefficients $C_i(x)$ reads: 
\ba
C_i&=& \sum_\q\hat C_i[\q](\as(m_Q)) \frac{\langle \ups\vert {\cal O}
(\q)\vert \ups\rangle}{m^{\delta_\q}}\, , \label{cexpt} 
\ea 
where $i=\gamma, q, {\overline q} ,g\;$  and
$\hat C_i[\q](x,\as(m_Q),)$  are the perturbative coefficients. 
The NRQCD sum is performed over all the relevant $\q$ states that
contribute at a desired order $v$.  The perturbative coefficients for
the colour octet states are now known at NLO~\cite{MP98}.  We can then
investigate the phenomenological applications of colour octet states,
if information on the NRQCD MEs can be obtained. To this aim we
proceed in two steps: first we obtain estimates by solving the
renormalization group (RG) equations~\cite{GK} and then we test the
reliability of these estimates analyzing their impact on the
observable $R_{\mu}(\Upsilon)=\Gamma(\ups\to{\rm
had})/\Gamma(\ups\to\mu^+\mu^-)$, whose value is well-known both
experimentally~\cite{PDG} and theoretically at NLO (both for singlet
and octet contributions~\cite{NLO}).  As a result the RG estimates
turn out to be smaller that one could expect form the na\"\i ve
velocity scaling rules and so legitimate doubts on their reliability
arise.  Nevertheless the above mentioned analysis shows that on one
hand RG estimates are to be thought as lower limit while on the other,
consistency between theory and experiment in total decay rates,
strongly disfavor much larger colour octet MEs.  I remark here that
these conclusions strongly rely on our present knowledge of the
theoretical expression which is only at NLO.  As a recent calculation
shows~\cite{NNLO}, NNLO corrections can be very large.  In
fig.~\ref{fig:plot-res2} are shown the various contributions to the
spectrum.  At low values of $z$ the fragmentation from octets is of
the same order of magnitude of the LO colour singlet one.  Contrary to
LO expectations in the framework of CSM~\cite{Catani}, we conclude
that the decay of $\Upsilon$ into a photon would not be useful for
estimating photon fragmentation functions.  For values of $z$ near the
end-point, breaking of the fixed-order calculation is manifest and the
resummation of both short-distance coefficient in $\as$ and the
non-perturbative MEs in v, is called for.  Finally we notice that the
overall effect of octet states is at its minimum in the central region
of the spectrum (exactly where the singlet LO direct contribution
dominates) and so it should be
used to make comparison with experimental data. Finally it has to be
stressed that relativistic and higher order strong corrections to the
singlet  should be included to have a consistent theoretical picture
at NLO\footnote{ The calculation of NLO corrections to the decay
$^3S_1^{[1]} \to gg\gamma$ are in progress~\cite{MK}.}.

\section{The inclusive decay $B \to J/\psi+ X$}
%
%
In the case of a inclusive B decay into a charmonium state
the following factorization formula holds~\cite{bbl}: 
\begin{equation}
\label{factform}
\Gamma(B\to H+X)=\sum_n C(b\to c\bar{c}[n]+x)\,\langle
{\cal O}^H[n]\rangle\,,
\end{equation}
which is valid up to power corrections of order $\Lambda_{QCD}/m_{b,c}$. 
(To this accuracy it is justified to treat the $B$ meson as a free $b$ quark.) 
The parameters $\langle{\cal O}^H[n]\rangle$, defined in \cite{bbl}, 
describe the hadronization of a couple of heavy quarks into a charmonium state
while the coefficient functions 
$C(b\to c\bar{c}[n]+x)$ describe the production of a $c\bar{c}$ 
configuration $n$ at short distances and can be expanded in the 
strong coupling $\alpha_s(\mu)$ at a scale $\mu$ of order $2 m_c$.
\noindent
The terms of interest in the $\Delta B = 1$ effective weak 
Hamiltonian\footnote{For simplicity 
I do not discuss here penguin contributions, although 
they have been included in the final results.} 
\ba  
H_{eff} = \frac{G_{F}}{\sqrt{2}} \sum_{q=s,d} 
V_{cb}^\ast V_{cq} \,
\left[ \frac{1}{3} C_{[1]} {\cal O}_1 + C_{[8]} 
{\cal O}_8 \right ]
\label{eq:Heff}  
\ea
contain the `current-current' operators   
\begin{eqnarray}
{\cal O}_1 &=&   
[\bar{c} \gamma_{\mu} (1-\gamma_5) c] \, 
[\bar{b} \gamma^{\mu} (1-\gamma_5) q]
\label{eq:opsing} \\  
{\cal O}_8 &=&    
[\bar{c}\,T^A \gamma_{\mu} (1-\gamma_5) c]\,  
[\bar{b}\,T^A \gamma^{\mu} (1-\gamma_5) q]\,. 
\label{eq:opoct}  
\end{eqnarray}
A next-to-leading order calculation has been recently
completed for all $S$ and $P$ states~\cite{BMR} but I  will focus 
here only on the decay into a $J/\psi$.
At leading order in the velocity expansion the 
spin-triplet $S$-wave charmonium states are produced directly from 
a $c\bar{c}$ pair with the same quantum numbers, i.e. 
$n={}^3\!S_1^{(1)}$. At order $v^4$ relative to this colour singlet 
contribution, a $\psi$ can materialize through the colour octet 
$c\bar{c}$ states $n={}^3\!S_1^{(8)}, {}^1\!S_0^{(8)}, 
{}^3\!P_J^{(8)}$, where the subscript `$J$' implies a sum over 
$J=0,1,2$. 
\begin{table}[t]
\setlength{\tabcolsep}{0.7pc}
\newlength{\digitwidth} \settowidth{\digitwidth}{\rm 0}
\catcode`?=\active \def?{\kern\digitwidth}
\caption{Extracted values for colour-octet MEs (in $10^{-2} {\rm GeV}^3$) for $J/\psi$ production  (numbers in parenthesis are input values).}
\label{tab:compare}
\begin{tabular}{lccc}
\hline
 Process  &$\langle {\cal O}_8^\psi (^3S_1) \rangle $& $k$ & ${\cal M}_k$\\
\hline
\hline
Tevatron\cite{BK97}             & $1.06  $ & $3.5 $  & $4.40  $ \\
f-t hadropr.\cite{BR96}         & $(0.66)$ & $ 7  $  & $3.0   $ \\
Tevatron\cite{CC97}             & $0.3   $ & $ 3  $  & $1.2   $ \\
Photoprod. \cite{AFM}           & $  -   $ & $ 7  $  & $2.0   $ \\
This work                       & $(1.06)$ & $3.1 $  & $1.5   $ \\
\hline
\hline
\end{tabular}
\end{table}
Note that even if the colour octet contributions are suppressed
by $v^4$ they must be included because 
the weak effective Hamiltonian favours the 
production of colour octet $c\bar{c}$ pairs. The approximate relation holds 
\begin{equation}
C_{[1]}^2 \sim  v^4 C_{[8]}^2,\,
\end{equation}
which shows how the NRQCD suppression is compensated by the structure
of the weak Hamiltonian.
\noindent 
In order to compare the final results with experimental data, we need 
some input information on MEs. 
We use $\langle {\cal O}_1^{\psi}({}^3\!S_1)\rangle=1.16\;\mbox{GeV}^3$
and the determination of $\langle {\cal O}_8^{\psi}({}^3\!S_1)
\rangle =1.06 \cdot 10^{-2}\,\mbox{GeV}^3$ 
from direct $J/\psi$ production at large 
transverse momentum in $p\bar{p}$ collisions \cite{BK97}.
The other two colour octet MEs are 
not yet well determined separately. 
Defining
\begin{equation}
M_{k}^\psi({}^1\!S_0^{(8)},{}^3\!P_J^{(8)})= 
\langle {\cal O}^{\psi}_8(^1\!S_0)  \rangle  +
\frac{k}{m_c^2}\,\langle {\cal O}^{\psi}_8(^3\!P_0) \rangle.
\end{equation}
we can reproduce the CLEO data $\mbox{Br}\,(B\to J/\psi+X)=(0.80\pm 0.08)\%$,
with :
\begin{equation}
\label{range}
M_{3.1}^\psi({}^1\!S_0^{(8)},{}^3\!P_J^{(8)})
=1.5_{-1.1}^{+0.8}\cdot 10^{-2}\,\mbox{GeV}^3\qquad 
\label{eq:final}
\end{equation}
where a conservative choice for the errors has been made.
It is interesting to compare the central values in (\ref{range}) 
and the upper limits with other determinations of the parameter 
$M_{k}^\psi({}^1\!S_0^{(8)},{}^3\!P_J^{(8)})$. 
This is summarized in Tab.~\ref{tab:compare}.
The central value in the first line is about a 
factor 3 larger than the central values obtained in~\ref{eq:final}.
As emphasized in Ref.~\cite{BK97} the 
Tevatron extraction is very sensitive to various effects that 
affect the transverse momentum distribution. Indeed, 
Refs.~\cite{CC97} quote smaller values compatible with, 
or smaller than the central values above. The total production 
cross section in fixed target collisions probes 
$M_{7}^\psi({}^1\!S_0^{(8)},{}^3\!P_J^{(8)})$ (assuming the 
validity of NRQCD factorization, which may be controversial). 
Given that a different combination of MEs enters, 
the values obtained in Ref.~\cite{BR96} are certainly consistent with 
the above central value. 
Considering the uncertainties involved in 
charmonium production in hadron collisions, 
the above upper limit on $M_{3.1}$
is the most stringent one existing at present. 

\section{Conclusions}
We have shown, by discussing both a decay and a production process, that the effects of including colour octet states contributions are relevant.
In particular we have given examples on how information on non-perturbative
MEs can be obtained by the comparison between theoretical quantities (where
short distance coefficients have been calculated at NLO) and available experimental
data.
This procedure has allowed us 
(a) to make predictions for photon energy spectrum in the $\Upsilon$ decays
gaining information from the total and leptonic decay rates;
(b) to give an estimate for some of colour octet MEs relevant for the $J/\psi$ production in B decays and so to check the ``universality'' by comparison with the values
extracted  from other processes.
As a final comment I would like to note that in both the above analysis we have found
values of the MEs {\it smaller } than one could expect from the na\" \i ve power
counting of NRQCD. Whether this is a general feature of NLO calculations (as a 
preliminary  analysis for fixed-target hadroproduction and photoproduction
indeed shows) and how this can be reconciled with the scaling rules is under study.

\section{Acknowledgements}
The results presented in this paper have been obtained in
collaboration with A.~Petrelli (Sec. 2) and with M.~Beneke,
I.~Rothstein (Sec. 3).  I am grateful to M.L.~Mangano for his valuable
advice and continuous support. 
Valuable discussions with S.~Catani, 
M.~Kr\"amer are gratefully acknowledged.  
I thank the CERN Theory Division for the hospitality while this work
was being carried out. 
Finally, I wish to thank Stephan Narison
for the financial help and the stimulating atmosphere at this
Conference. 
 \\
This work was supported by the EU Fourth Framework
Programme `Training and Mobility of Researchers', Network `Quantum
Chromodynamics and the Deep Structure of Elementary Particles',
contract FMRX-CT98-0194 (DG 12 - MIHT).

\end{document}